\documentclass[final]{svjour3_compact_authors}
\usepackage[]{graphicx}
\usepackage{rotating}
\usepackage{amssymb}
\usepackage{mathptmx}
\usepackage{xcolor}
\usepackage{siunitx}

\DeclareGraphicsExtensions{.pdf,.png,.jpg}

\usepackage[
    backend=bibtex,
    natbib=true,
    url=false, 
    doi=true,
    eprint=false,
    giveninits=true,
    maxnames=1,
    sorting=none
]{biblatex}
\AtEveryBibitem{\clearfield{title}}
\renewbibmacro{in:}{}
\DefineBibliographyStrings{english}{%
  page             = {\ifbibliography{}{\adddot}},
  pages            = {\ifbibliography{}{\adddot}},
}
\DeclareFieldFormat{postnote}{#1}
\usepackage{array}
\newcolumntype{L}{>{\centering\arraybackslash}m{1.55cm}}

\makeatletter
\journalname{Journal of Low Temperature Physics}

\begin{document}

\newcommand{\hdblarrow}{H\makebox[0.9ex][l]{$\downdownarrows$}-}
\title{On-sky performance of the SPT-3G frequency-domain multiplexed readout}

\author{
A.~N.~Bender\textsuperscript{1,2} \and 
A.~J.~Anderson\textsuperscript{3,2} \and 
J.~S.~Avva\textsuperscript{4} \and 
P.~A.~R.~Ade\textsuperscript{5} \and 
Z.~Ahmed\textsuperscript{6,7} \and 
P.~S.~Barry\textsuperscript{1,2} \and 
R.~Basu Thakur\textsuperscript{2} \and 
B.~A.~Benson\textsuperscript{3,2,8} \and 
L.~Bryant\textsuperscript{9} \and 
K.~Byrum\textsuperscript{1} \and 
J.~E.~Carlstrom\textsuperscript{2,9,10,1,8} \and 
F.~W.~Carter\textsuperscript{1,2} \and 
T.~W.~Cecil\textsuperscript{1} \and 
C.~L.~Chang\textsuperscript{1,2,8} \and 
H.-M.~Cho\textsuperscript{7} \and 
J.~F.~Cliche\textsuperscript{11} \and 
A.~Cukierman\textsuperscript{4} \and 
T.~de~Haan\textsuperscript{4} \and 
E.~V.~Denison\textsuperscript{12} \and 
J.~Ding\textsuperscript{13} \and 
M.~A.~Dobbs\textsuperscript{11,14} \and 
D.~Dutcher\textsuperscript{2,10} \and 
W.~Everett\textsuperscript{15} \and 
K.~R.~Ferguson\textsuperscript{16} \and 
A.~Foster\textsuperscript{17} \and 
J.~Fu\textsuperscript{18} \and 
J.~Gallicchio\textsuperscript{2,19} \and 
A.~E.~Gambrel\textsuperscript{2} \and 
R.~W.~Gardner\textsuperscript{9} \and 
A.~Gilbert\textsuperscript{11} \and 
J.~C.~Groh\textsuperscript{4} \and 
S.~Guns\textsuperscript{4} \and 
R.~Guyser\textsuperscript{18} \and 
N.~W.~Halverson\textsuperscript{15,20} \and 
A.~H.~Harke-Hosemann\textsuperscript{1,18} \and 
N.~L.~Harrington\textsuperscript{4} \and 
J.~W.~Henning\textsuperscript{1,2} \and 
G.~C.~Hilton\textsuperscript{12} \and 
W.~L.~Holzapfel\textsuperscript{4} \and 
D.~Howe\textsuperscript{21} \and 
N.~Huang\textsuperscript{4} \and 
K.~D.~Irwin\textsuperscript{6,22,7} \and 
O.~B.~Jeong\textsuperscript{4} \and 
M.~Jonas\textsuperscript{3} \and 
A.~Jones\textsuperscript{21} \and 
T.~S.~Khaire\textsuperscript{13} \and 
A.~M.~Kofman\textsuperscript{18} \and 
M.~Korman\textsuperscript{17} \and 
D.~L.~Kubik\textsuperscript{3} \and 
S.~Kuhlmann\textsuperscript{1} \and 
C.-L.~Kuo\textsuperscript{6,22,7} \and 
A.~T.~Lee\textsuperscript{4,23} \and 
E.~M.~Leitch\textsuperscript{2,8} \and 
A.~E.~Lowitz\textsuperscript{2} \and 
S.~S.~Meyer\textsuperscript{2,9,10,8} \and 
D.~Michalik\textsuperscript{21} \and 
J.~Montgomery\textsuperscript{11} \and 
A.~Nadolski\textsuperscript{18} \and 
T.~Natoli\textsuperscript{24} \and 
H.~Nguyen\textsuperscript{3} \and 
G.~I.~Noble\textsuperscript{11} \and 
V.~Novosad\textsuperscript{13} \and 
S.~Padin\textsuperscript{2} \and 
Z.~Pan\textsuperscript{2,10} \and 
P.~Paschos\textsuperscript{9} \and 
J.~Pearson\textsuperscript{13} \and 
C.~M.~Posada\textsuperscript{13} \and 
W.~Quan\textsuperscript{2,10} \and 
A.~Rahlin\textsuperscript{3,2} \and 
D.~Riebel\textsuperscript{21} \and 
J.~E.~Ruhl\textsuperscript{17} \and 
J.T.~Sayre\textsuperscript{15} \and 
E.~Shirokoff\textsuperscript{2,8} \and 
G.~Smecher\textsuperscript{25} \and 
J.~A.~Sobrin\textsuperscript{2,10} \and 
A.~A.~Stark\textsuperscript{26} \and 
J.~Stephen\textsuperscript{9} \and 
K.~T.~Story\textsuperscript{6,22} \and 
A.~Suzuki\textsuperscript{23} \and 
K.~L.~Thompson\textsuperscript{6,22,7} \and 
C.~Tucker\textsuperscript{5} \and 
L.~R.~Vale\textsuperscript{12} \and 
K.~Vanderlinde\textsuperscript{24,27} \and 
J.~D.~Vieira\textsuperscript{18,28} \and 
G.~Wang\textsuperscript{1} \and 
N.~Whitehorn\textsuperscript{16} \and 
V.~Yefremenko\textsuperscript{1} \and 
K.~W.~Yoon\textsuperscript{6,22,7} \and 
M.~R.~Young\textsuperscript{27} \and 
}
\institute{
\textsuperscript{1}{High-Energy Physics Division, Argonne National Laboratory, 9700 South Cass Avenue., Argonne, IL, 60439, USA} \\
\textsuperscript{2}{Kavli Institute for Cosmological Physics, University of Chicago, 5640 South Ellis Avenue, Chicago, IL, 60637, USA} \\
\textsuperscript{3}{Fermi National Accelerator Laboratory, MS209, P.O. Box 500, Batavia, IL, 60510, USA} \\
\textsuperscript{4}{Department of Physics, University of California, Berkeley, CA, 94720, USA} \\
\textsuperscript{5}{School of Physics and Astronomy, Cardiff University, Cardiff CF24 3YB, United Kingdom} \\
\textsuperscript{6}{Kavli Institute for Particle Astrophysics and Cosmology, Stanford University, 452 Lomita Mall, Stanford, CA, 94305, USA} \\
\textsuperscript{7}{SLAC National Accelerator Laboratory, 2575 Sand Hill Road, Menlo Park, CA, 94025, USA} \\
\textsuperscript{8}{Department of Astronomy and Astrophysics, University of Chicago, 5640 South Ellis Avenue, Chicago, IL, 60637, USA} \\
\textsuperscript{9}{Enrico Fermi Institute, University of Chicago, 5640 South Ellis Avenue, Chicago, IL, 60637, USA} \\
\textsuperscript{10}{Department of Physics, University of Chicago, 5640 South Ellis Avenue, Chicago, IL, 60637, USA} \\
\textsuperscript{11}{Department of Physics and McGill Space Institute, McGill University, 3600 Rue University, Montreal, Quebec H3A 2T8, Canada} \\
\textsuperscript{12}{NIST Quantum Devices Group, 325 Broadway Mailcode 817.03, Boulder, CO, 80305, USA} \\
\textsuperscript{13}{Materials Sciences Division, Argonne National Laboratory, 9700 South Cass Avenue, Argonne, IL, 60439, USA} \\
\textsuperscript{14}{Canadian Institute for Advanced Research, CIFAR Program in Gravity and the Extreme Universe, Toronto, ON, M5G 1Z8, Canada} \\
\textsuperscript{15}{CASA, Department of Astrophysical and Planetary Sciences, University of Colorado, Boulder, CO, 80309, USA } \\
\textsuperscript{16}{Department of Physics and Astronomy, University of California, Los Angeles, CA, 90095, USA} \\
\textsuperscript{17}{Department of Physics, Center for Education and Research in Cosmology and Astrophysics, Case Western Reserve University, Cleveland, OH, 44106, USA} \\
\textsuperscript{18}{Department of Astronomy, University of Illinois at Urbana-Champaign, 1002 West Green Street, Urbana, IL, 61801, USA} \\
\textsuperscript{19}{Harvey Mudd College, 301 Platt Boulevard., Claremont, CA, 91711, USA} \\
\textsuperscript{20}{Department of Physics, University of Colorado, Boulder, CO, 80309, USA} \\
\textsuperscript{21}{University of Chicago, 5640 South Ellis Avenue, Chicago, IL, 60637, USA} \\
\textsuperscript{22}{Deptartment of Physics, Stanford University, 382 Via Pueblo Mall, Stanford, CA, 94305, USA} \\
\textsuperscript{23}{Physics Division, Lawrence Berkeley National Laboratory, Berkeley, CA, 94720, USA} \\
\textsuperscript{24}{Dunlap Institute for Astronomy \& Astrophysics, University of Toronto, 50 St. George Street, Toronto, ON, M5S 3H4, Canada} \\
\textsuperscript{25}{Three-Speed Logic, Inc., Vancouver, B.C., V6A 2J8, Canada} \\
\textsuperscript{26}{Harvard-Smithsonian Center for Astrophysics, 60 Garden Street, Cambridge, MA, 02138, USA} \\
\textsuperscript{27}{Department of Astronomy \& Astrophysics, University of Toronto, 50 St. George Street, Toronto, ON, M5S 3H4, Canada} \\
\textsuperscript{28}{Department of Physics, University of Illinois Urbana-Champaign, 1110 West Green Street, Urbana, IL, 61801, USA} \\
}

\maketitle

\begin{abstract}

Frequency-domain multiplexing (fMux) is an established technique for the readout of large arrays of transition edge sensor (TES) bolometers.  Each TES in a multiplexing module has a unique AC voltage bias that is selected by a resonant filter. This scheme enables the operation and readout of multiple bolometers on a single pair of wires, reducing thermal loading onto sub-Kelvin stages.  The current receiver on the South Pole Telescope, SPT-3G, uses a 68x fMux system to operate its large-format camera of $\sim$16,000 TES bolometers.  
We present here the successful implementation and performance of the SPT-3G  readout as measured on-sky.   Characterization of the noise reveals a median pair-differenced 1/f knee frequency of 33 mHz, indicating that low-frequency noise in the readout will not limit SPT-3G's measurements of sky power on large angular scales.  Measurements also show that the median readout white noise level in each of the SPT-3G observing bands is below the expectation for photon noise, demonstrating that SPT-3G is operating in the photon-noise-dominated regime. 


\end{abstract}

\section{Introduction}
Multiplexing readout is an enabling technology for current and future cosmic microwave background (CMB) receivers.  Large arrays of superconducting detectors require multiplexing readout to reduce the number of wires that provide electrical connections to  the detectors.  Fewer wires dissipate less heat on the sub-Kelvin stage and reduce the complexity of the readout, which can result in lower overall cost. 
Several different multiplexing schemes  exist including time-division multiplexing \cite{dekorte2003,battistelli2008}, megahertz frequency-domain multiplexing \cite{dobbs2012,gottardi2016}, microwave multiplexing \cite{dober2018}, and code-division multiplexing \cite{stiehl2012}.  Modern CMB receivers have or are working to implement the first three schemes, each of which brings a unique set of advantages and challenges \cite{Ade_2015, henderson2016,suzuki2016, bender2018, galitzki2018}. 

The SPT-3G receiver uses frequency-domain multiplexing (fMux) to read
out its focal plane of $\sim$16,000 transition-edge sensor (TES) bolometers.
Installed on the South Pole Telescope in late 2016, SPT-3G is
currently making deep, high-resolution maps of the CMB temperature and polarization.
These data will be used to constrain structure formation in the  late
universe and search for the unique signature of inflation in the early
universe.   The successful implementation of fMux readout in SPT-3G was a key technological advancement that has enabled this large-format array to achieve background-limited performance.  
In this paper we present the performance of the SPT-3G fMux readout, measured during CMB survey operations.  Sec. \ref{sec:implementation} describes the specific implementation of fMux in SPT-3G.  Sec. \ref{sec:performance}  discusses the performance of the readout, including crosstalk and noise measurements. 

\vspace{-14pt}
\section{SPT-3G Readout Implementation}
\label{sec:implementation}

\begin{figure}[t]
\begin{center}
\includegraphics[width=0.90\linewidth,keepaspectratio]{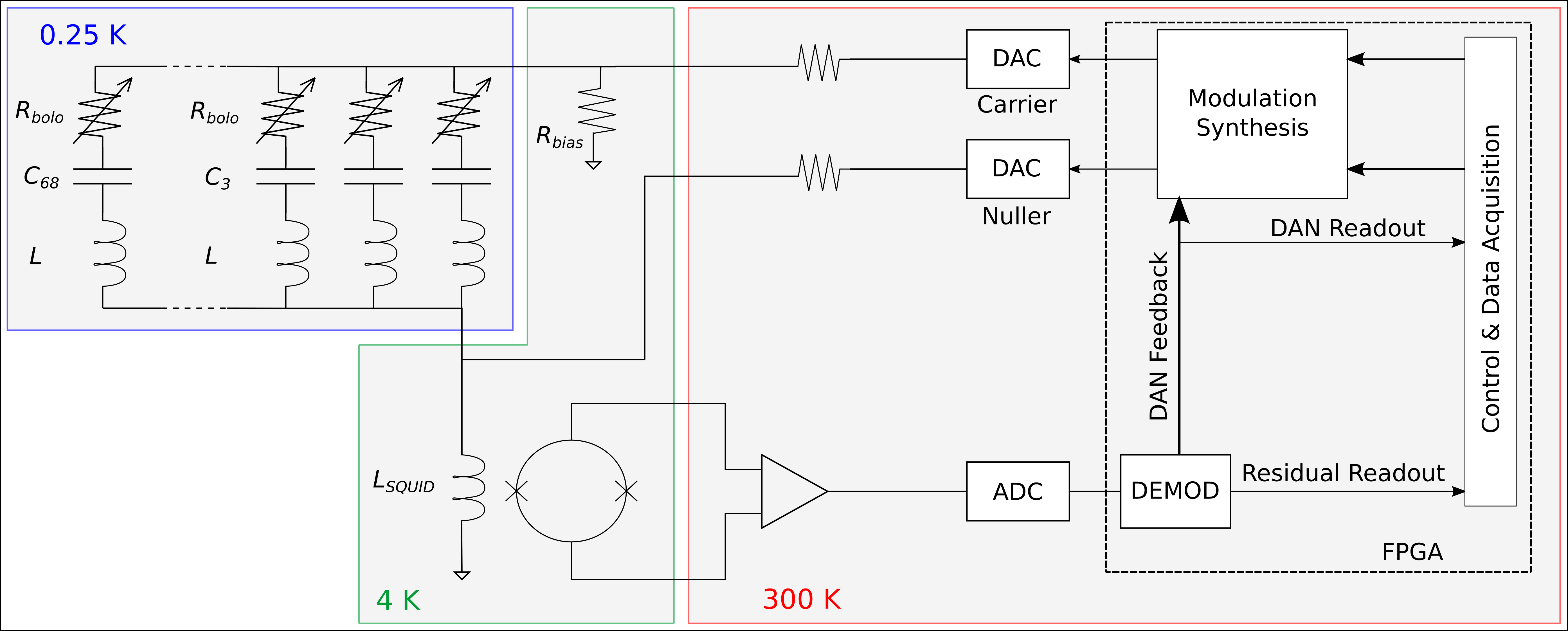}
\caption{A schematic diagram of the SPT-3G implementation of the frequency-domain multiplexed readout.  The different colors on the diagram represent the different operating temperatures of components in the system. (Color figure online).}
\label{fig:readoutschematic}
\end{center}
\vspace{-6mm}
\end{figure}

A schematic representation of the SPT-3G fMux readout is shown in Fig. \ref{fig:readoutschematic}.  In fMux readout, each TES is connected in series with an inductor and capacitor, creating a unique resonant bandwidth.    Segments are connected in parallel, allowing a waveform consisting of summed sinusoidal voltage biases (shown as the carrier in Fig. \ref{fig:readoutschematic})  to be input and the subsequent signals read out on a single pair of wires.  The $LC$ filter selects the appropriate AC sinusoid from the waveform to bias the associated TES.  The ability to provide individually optimized biases to every bolometer is a unique advantageous feature of fMux readout.  SPT-3G combines 66 TES bolometers and 2 calibration resistors together into a single module for a total multiplexing factor of 68.   

In the limit of high electro-thermal loop gain, the sum of electrical and optical power on the detector is fixed. The TES is biased with a constant amplitude voltage. Changes in optical power amplitude modulate the current flowing through the TES, encoding the signals of interest in the sidebands. 
The currents flowing through the TES bolometers are summed together and amplified by a superconducting quantum interference device (SQUID). After further amplification the output is digitized and demodulated.    

In order to reduce the dynamic range requirement on the SQUID, a nulling current is injected into the SQUID input.  The nuller removes both the AC bias waveform and provides continuously updating feedback to remove the sidebands (digital active nulling) \cite{dehaan2012}.

\begin{figure}[t]
\begin{center}
\includegraphics[height=0.19\textheight,keepaspectratio]{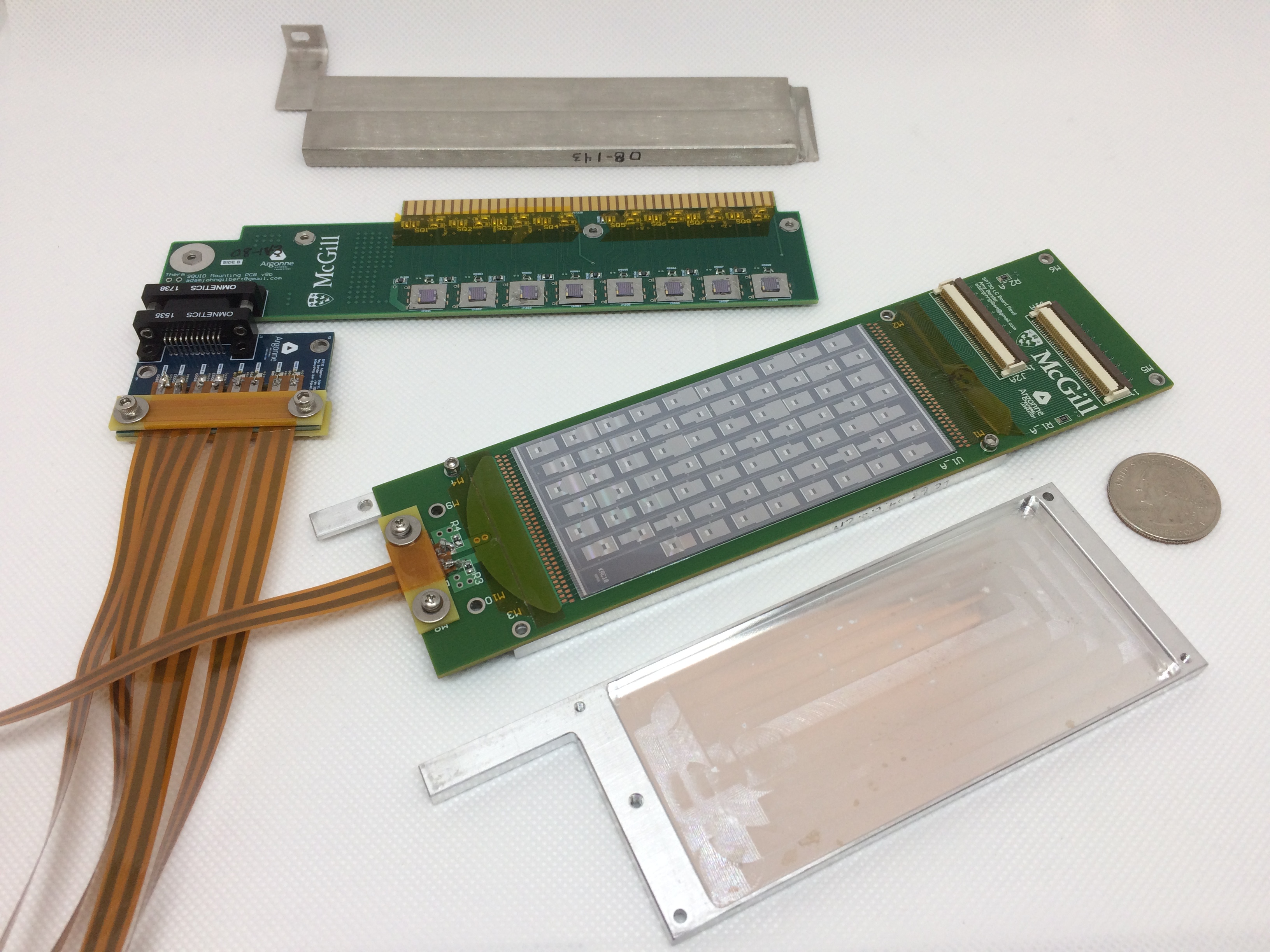}
\includegraphics[height=0.19\textheight,keepaspectratio]{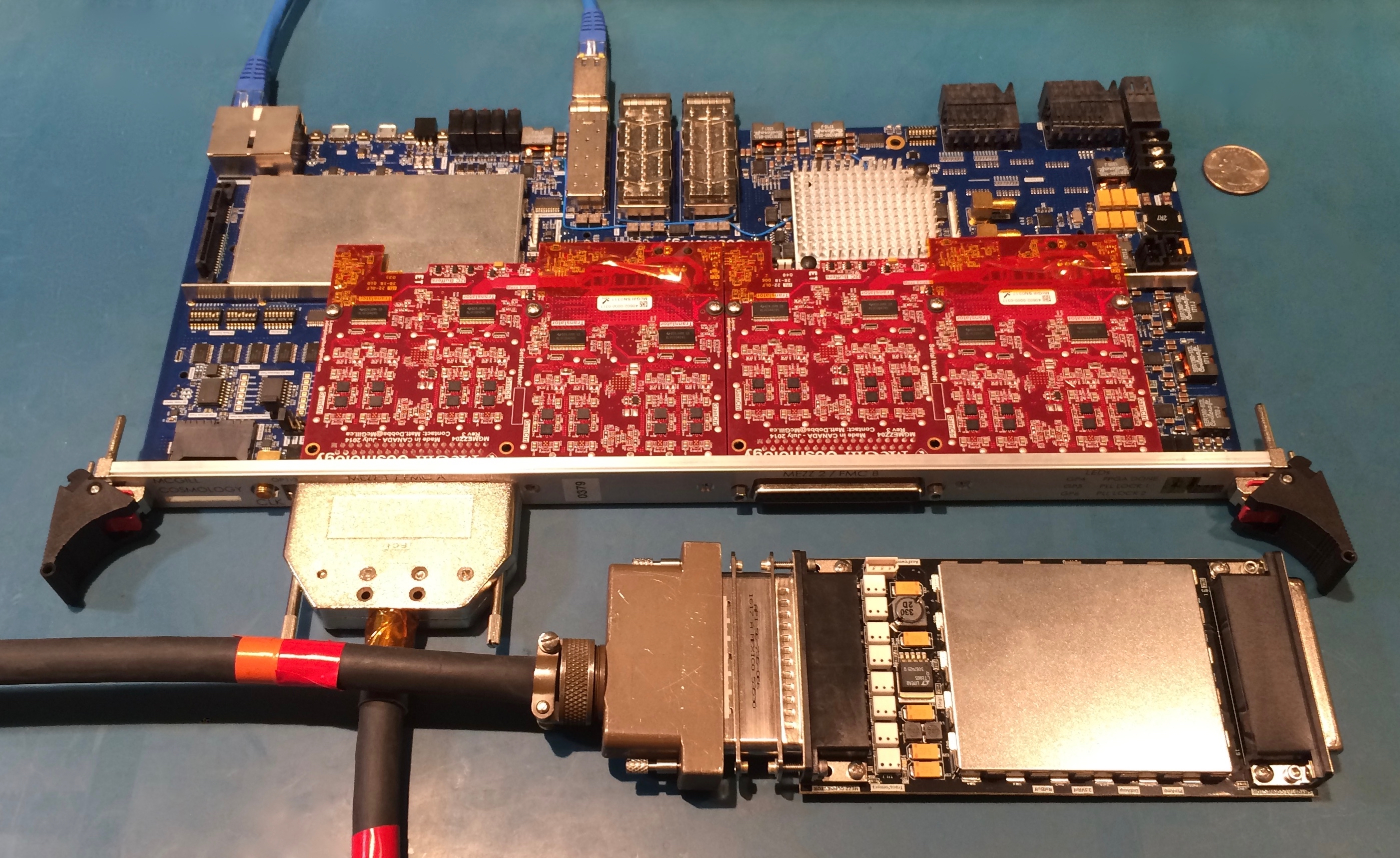}
\caption{\textit{Left:} The cryogenic components in the SPT-3G fMux readout. In the center of the photograph is a single 68x multiplexing $LC$ filter network chip that connects to the SQUID (top of photograph) via stripline wiring.  Also shown is the mechanical shield for the $LC$ chip and magnetic shield for the SQUIDs. \textit{Right:} The custom FPGA and SQUID control boards. (Color figure online.)}
\label{fig:cryoreadoutphoto}
\end{center}
\vspace{-5mm}
\end{figure}

The left-hand image in Fig. \ref{fig:cryoreadoutphoto} shows the cryogenic readout components in the SPT-3G fMux.  The $LC$ network for each module of 68 channels consists of a chip with pairs of superconducting inductors and capacitors that is mounted on a standard printed circuit board \cite{hattori2014, rotermund2016,bender2016}.  A superconducting broadside-coupled stripline connects the $LC$ assemblies to the SQUIDs, which are mounted and shielded in groups of eight on printed circuit boards \cite{avva2018}.  The right-hand image in Fig. \ref{fig:cryoreadoutphoto} shows the room temperature electronics.  Custom FPGA boards perform the required digital signal processing (modulation, demodulation, and feedback).   A second custom board operates the SQUIDs \cite{bender2014,bandura2016}.

\vspace{-10pt}

\section{On-Sky Performance}
\label{sec:performance}

\subsection{Yield}
The overall sensitivity of a CMB receiver scales directly with the number of optically sensitive detectors. Any readout components that are inoperable reduce the total number of useable detectors.  The total number of resonant filters identified in the characterization of the SPT-3G readout combines detector and readout yield together.  The detector wafers are electrically probed at room temperature prior to assembly, providing an independent estimate of detector yield.  The table below presents a summary of this yield accounting.   The readout yield is estimated from the number of identified resonances, accounting for detectors that measured disconnected in a room temperature probing.
Loss in the readout occurs in the cryogenic components in two main ways.  First, an entire module is disconnected either at the stripline connections or the $LC$ chip.  Current data suggests this has occurred for 2-3 out of the 240 modules within SPT-3G.  Second, individual $LC$ pairs are non-functional, accounting for the remainder of missing resonances.
\begin{table}[h]
\begin{center}
\begin{tabular}{L|L|L|L|L|L}
Wired Readout Channels  &Calibration Resistors & TES with T=300K connectivity& Expected Resonances &Identified Resonances  &Estimated Readout Yield \\
\hline
\vspace{5pt}
16,200 & 480 & 14,712& 15,192 & 14,260& $\sim$94\%\\
\end{tabular}
\end{center}
\vspace{-10mm}
\end{table}

\subsection{Crosstalk}
Electrical crosstalk from the fMux system copies sky signals from one detector into the data of another. 
The resulting bias on the measured CMB polarization is either a leakage of temperature signal into polarization or an incorrect polarized beam \cite{Crites_2015,kiesler2015}.  There are three expected types of crosstalk within an fMux module \cite{dobbs2012}.  First, there is crosstalk due to overlapping bandwidth between $LC$ filters closely spaced in frequency.  Second, the inductors within a module can couple via mutual inductance.  Finally, the stripline wiring that connects the filter and TES network to the SQUID has an impedance.  This impedance acts as a voltage divider, creating a mechanism to modulate the AC biases slightly as TES resistance responds to sky signal.  

\begin{figure}[t]
\begin{center}
\includegraphics[width=0.49\linewidth,keepaspectratio]{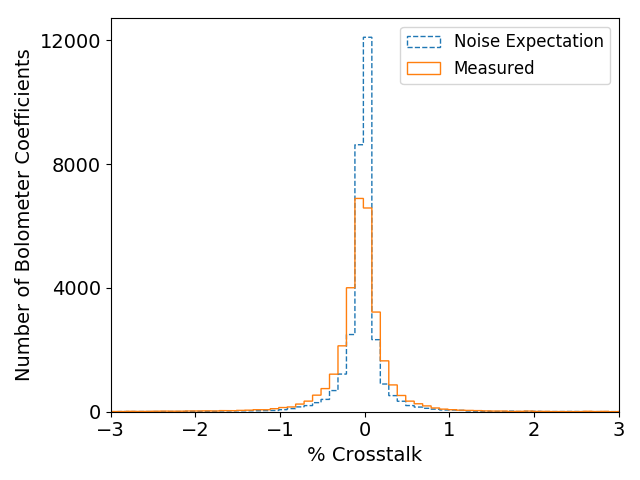}
\includegraphics[width=0.49\linewidth,keepaspectratio]{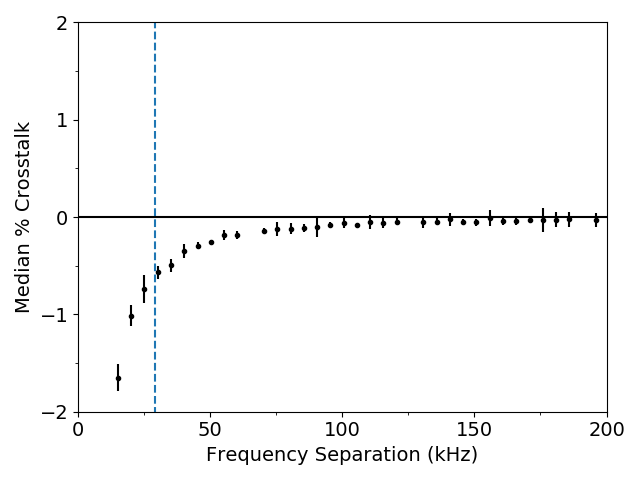}
\caption{\textit{Left:} The measured distribution of crosstalk coefficients in the SPT-3G receiver (\textit{orange line}).  The \textit{blue dashed} line shows the noise expectation for the measurement. \textit{Right:}The dependence of the crosstalk coefficients on the spacing between the two AC bias frequencies.  The points are the median value across 14 observations with the error bars showing the uncertainty in the mean of the distribution. The \textit{blue dashed} line represents the designed minimum frequency spacing that predicted crosstalk coefficients $<0.5$\%. (Color figure online.)}
\label{fig:xtalk}
\end{center}
\vspace{-5mm}
\end{figure}

Crosstalk is measured for SPT-3G using observations of the galactic HII region RCW38.  SPT-3G regularly observes RCW38 as part of its calibration scheme, raster scanning the telescope such that every detector in the focal plane sees the source.   RCW38 is slightly extended compared to a true point source.  Therefore, a template for the expected flux distribution is constructed from the average for all detectors within a given observing band (95/150/220 GHz) and detector wafer.  A map of RCW38 is then made for each detector individually and the associated template is used to measure the flux at the expected source position (the primary signal from that detector).  Flux is also measured at the known offsets for all detectors within the same readout module.  The ratio of these two flux measurements quantifies the crosstalk coefficients for each detector.    The procedure is also applied to a region of the map outside the source region to generate a noise expectation.  Fig. \ref{fig:xtalk} shows the resulting crosstalk distribution and the noise expectation.

There are several subtleties present in the crosstalk analysis.  First, this technique measures the total crosstalk in SPT-3G, including any optical contribution.  However, negative crosstalk coefficients can only originate in the readout.  Next, this analysis is unable to separate the individual crosstalk coefficients from each of  the bolometers within the same pixel.  Consequently, the crosstalk coefficient is assigned to the detector with the bias frequency closest to that of the primary detector.   As mentioned previously, RCW38 is slightly extended. Spatial neighbors in the focal plane often fall within the source envelope, confusing the crosstalk measurement.  There is no correlation between focal plane position and relative spacing of bias frequencies in SPT-3G.  Therefore, detector pairs separated by less than 20 arcminutes are excluded in this characterization of readout crosstalk.  We note a small excess of positive crosstalk in Fig. \ref{fig:xtalk}, which could be residual from this effect or non-readout crosstalk. Finally, as described previously, crosstalk coefficients are only calculated between detectors within the same readout module.  When the measurement is expanded to include wafer-level crosstalk coefficients, there is no additional crosstalk significant in comparison to the noise expectation. 

The cryogenic readout components for SPT-3G were designed to maintain crosstalk at a level of 0.5\% or below, assuming zero scatter in the resonant frequencies \cite{bender2016}.  There is a tail in the distribution of crosstalk components shown in Fig. \ref{fig:xtalk} that exceeds this specification.  To understand this deviation, the median crosstalk level across multiple RCW38 observations is plotted as a function of the bias frequency separation between the detectors (shown in the right-hand panel of Fig. \ref{fig:xtalk}).   The vertical dashed line denotes the minimum frequency spacing in the design.  The crosstalk coefficients in excess of 0.5\% clearly correlate with bias frequencies that are closer together than designed (a result of small deviations from design in the fabricated capacitors).   Future fMux systems with improved control of this frequency scatter or other design changes that mitigate the mechanisms by which crosstalk depends on bias frequency spacing will suppress this excess. 

\vspace{-5mm}
\subsection{Noise}

A critical metric of performance for the SPT-3G receiver is the noise.  The total noise during observations is a combination of photon noise from the sky and optical elements in the receiver itself, phonon  and Johnson noise intrinsic to the bolometers, and finally readout noise.   To measure the readout noise, the TES bolometers are first operated into the transition with the telescope at a typical observing elevation. This strategy ensures that the bias and nuller waveforms approximate nominal levels during the noise measurement.  The telescope then moves to point at the horizon, increasing the optical loading enough to saturate the bolometers and suppress the photon and phonon noise.   Johnson noise will still be present at the level of $\sim2.6$ pA/$\sqrt{Hz}$ for a 2\,\si{\ohm} TES operating at a temperature of 315 mK, but the resulting data will be dominated by the readout noise.  Data is acquired for five minutes while the telescope is held stationary (referred to as a noise stare).

\begin{figure}[t]
\begin{center}
\includegraphics[width=0.995\textwidth,keepaspectratio]{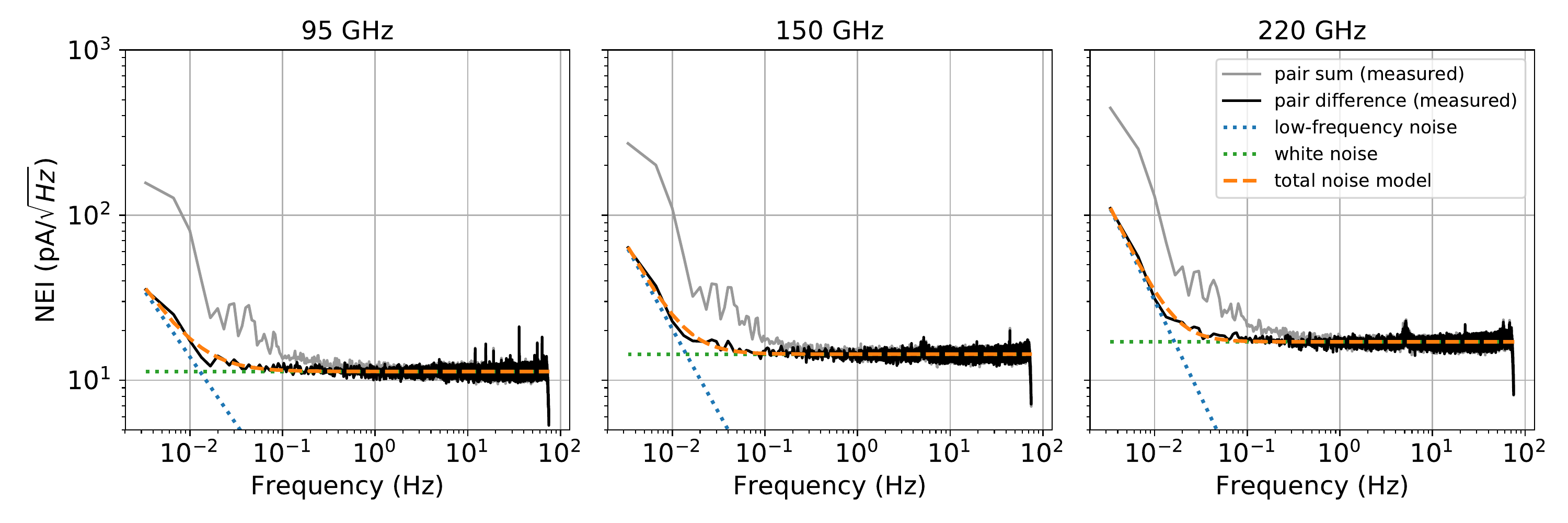}
\caption{Average pair-sum (\textit{light grey}) and pair-difference (\textit{black}) amplitude spectral distributions for a single wafer in the SPT-3G focal plane.  These data are taken with the detectors saturated, resulting in a measurement of the fMux noise behavior. The\textit{ blue} and \textit{green dotted} lines show the 1/f and white noise components of the model fit. The \textit{orange dashed} line shows the best-fit to the total noise model. (Color figure online.) }
\label{fig:horizonnoisepsd}
\end{center}
\vspace{-5mm}
\end{figure}

\begin{figure}[t]
\begin{center}
\includegraphics[height=0.21\textheight,keepaspectratio]{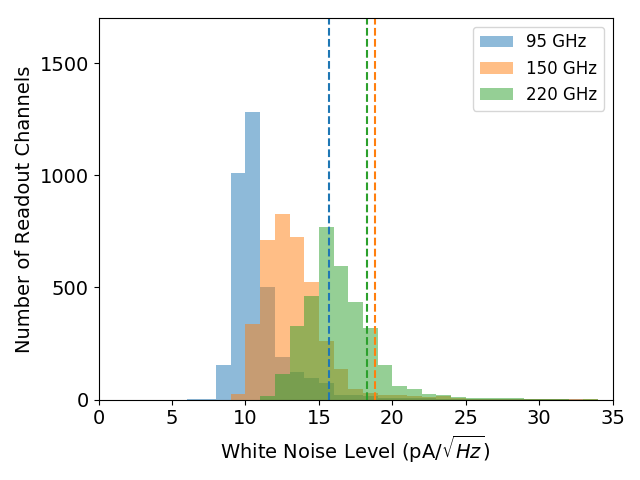}
\includegraphics[height=0.21\textheight,keepaspectratio]{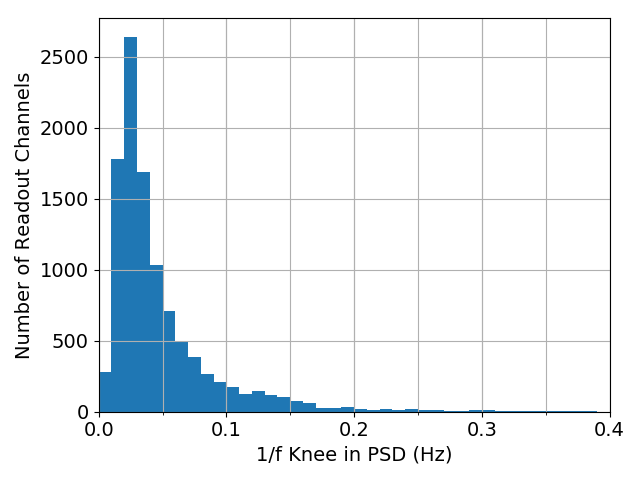}
\caption{\textit{Left:} The distribution of white noise measured in the SPT-3G fMux readout. The \textit{dashed} lines depict the expected photon noise level for each of the observing bands.  Note that 220 GHz photon noise is less than the 150 GHz due to a combination of decreased optical efficiency and increased voltage bias. \textit{Right:} The distribution of 1/f knee frequencies from the individual pair-difference readout noise spectra. (Color figure online.)}
\label{fig:horizonnoisehist}
\end{center} 
\end{figure}

Amplitude spectral distributions (ASDs) are extracted from the noise stare by taking the sum and difference of time-ordered data (TOD) for detector pairs of the same observing band within each pixel.   The TOD are initially  conditioned by removing a first-order polynomial.  The TOD are then gain-matched in the low-frequency region ($0.01 < f < 1$ Hz) to remove correlated noise between the detector pairs and measure the fundamental low-frequency noise behavior.  Fig. \ref{fig:horizonnoisepsd} shows representative pair-sum and pair-difference ASDs averaged for all detectors within an observing band on a single detector wafer in the SPT-3G focal plane.  A noise model containing white noise and 1/f noise  (NEI$_{\textrm{total}}$ =  $\sqrt{B + A\cdot f^{-\alpha}}$) is fit to the difference ASD for each detector pair.  The 1/f knee is derived from the resulting fit parameters by calculating the frequency $f_c$ at which the 1/f noise equals the white noise level in the ASD.  Histograms of the readout white noise level and 1/f knee are shown in Fig. \ref{fig:horizonnoisehist} and median values are presented in the table below.

Several important features of the SPT-3G fMux readout are observed in these results.  First, the pair-sum ASDs show increased noise at low frequency compared to the pair-difference.  This excess implies correlated low-frequency noise that originates within the readout via a mechanism that affects a module as whole.   Expected sources include temperature fluctuations in electronics (both cryogenic and room temperature) and noise intrinsic to the digital-to-analog converters.  It is important to note that multiple analysis techniques exist for cleaning correlated noise from data (including pair-differencing), leaving only the contribution from the uncorrelated low-frequency readout noise.

\vspace{-4mm}
\begin{table}[h]
\begin{center}
\begin{tabular}{c|c|c}
Observing Band & White Noise Level & 1/f Frequency  \\
{[GHz]} & pA/{[$\sqrt{Hz}$]} & {[Hz]} \\
\hline
95 GHz & 10.4&0.041\\
150 GHz &13.0&0.032\\
220 GHz &16.0&0.029\\
\end{tabular}
\end{center}
\vspace{-9mm}
\end{table}

The next feature evident in both the ASDs and Fig. \ref{fig:horizonnoisehist} is that the white noise level appears to increase across the 95/150/220 GHz bands.   In SPT-3G, the different bands are grouped together across the electronics bandwidth (i.e., the 95 GHz detectors have the lowest AC bias frequencies, followed by the 150 GHz and 220 GHz).  The white noise level is therefore increasing with bias frequency (and not observing band), resulting from a subtle feature in the nulling scheme.  Briefly, a small fraction of the nuller leaks back through the comb and bias resistor to ground and the nuller increases to compensate, amplifying noise sources in the demodulation chain \cite{bender2018}.  The amplification factor depends on the ratio of comb impedance and squid input-coil impedance. Future implementations of fMux can mitigate this leakage effect by changing the impedance ratio.  The histogram of white noise shown in Fig. \ref{fig:horizonnoisehist} also includes the expected photon noise level for the three SPT-3G observing bands. The median readout white noise level is clearly below the photon noise, showing that despite the frequency dependence, the SPT-3G fMux readout is operating as desired in the photon-noise-dominated regime. 

Finally, we see in Fig. \ref{fig:horizonnoisehist} that the median 1/f knee frequency in the pair-differenced ASDs is $f_{c}\sim33\:$mHz.  Typical CMB telescopes scan the sky at a rate of $\sim 0.5^{\circ}/s$ or faster, up-mixing the science band to separate it from low-frequency noise sources.  With this scan speed, the fMux readout 1/f knee $f_c$ corresponds to a multipole number of $\ell \sim 24$.  At multipoles below this (larger angular scales on the sky), there will be increasing levels of readout noise.  This knee is around a factor of three lower in multipole than the peak of predicted inflationary CMB polarization spectrum from recombination, a key scientific target for modern CMB experiments \cite{kamionkowski2016}. Additionally, both pair-sum (CMB temperature) and pair-difference (CMB polarization) measurements will have low-frequency noise contributions from the atmosphere.  Even at an excellent site such as the South Pole, the low-frequency atmospheric contribution is expected to dominate the readout contribution presented here \cite{bussmann2005}.  We observe that low-frequency noise from the fMux readout does not limit SPT-3G measurements of either polarization or temperature power at the scales relevant to the key science goals.

\vspace{-5mm}
\section{Summary}

We present the on-sky performance of the SPT-3G fMux readout.  Crosstalk is shown to meet the design specification of $<0.5\%$, with a slight excess resulting from increased frequency scatter in the $LC$ filter network.  We measure the white noise level of the readout and demonstrate that SPT-3G is operating in the photon-noise-dominated regime.  Additionally, we explore the low-frequency noise contribution from the readout and measure a median readout 1/f knee frequency of 33 mHz.  Low-frequency noise from the fMux readout will therefore  not limit the SPT-3G receiver's ability to measure the CMB polarization at the large angular scales of the inflationary signal from recombination.

\begin{acknowledgements}
The South Pole Telescope program is supported by the National Science
Foundation (NSF)  through grant PLR-1248097. Partial support is also provided by the NSF Physics Frontier Center grant PHY-1125897 to the Kavli Institute of Cosmological Physics at the University of Chicago, the Kavli Foundation, and the Gordon and Betty Moore Foundation through grant GBMF\#947 to the University of Chicago.  Work at Argonne National Lab is supported by UChicago Argonne LLC, Operator of Argonne National Laboratory (Argonne). Argonne, a U.S. Department of Energy Office of Science Laboratory, is operated under contract no. DE-AC02-06CH11357. We acknowledge R. Divan, L. Stan, C.S. Miller, and V. Kutepova for supporting our work in the Argonne Center for Nanoscale Materials.
Work at Fermi National Accelerator Laboratory, a DOE-OS, HEP User Facility managed by the Fermi Research Alliance, LLC, was supported under Contract No. DE-AC02-07CH11359.  NWH acknowledges support from NSF CAREER grant AST-0956135. The McGill authors acknowledge funding from the Natural Sciences and Engineering Research Council of Canada, Canadian Institute for Advanced Research, and the Fonds de recherche du Québec Nature et technologies. JV acknowledges support from the Sloan Foundation.
\end{acknowledgements}

\printbibliography

\end{document}